\documentclass[showkeys,aps,prl,reprint,superscriptaddress,amsmath,amssymb]{revtex4-1}
\usepackage{graphicx,color}

\begin{document}
\title{Connecting water correlations, fluctuations, and wetting phenomena at hydrophobic and hydrophilic surfaces}

\author{Rahul Godawat}
\thanks{RG and SNJ contributed equally to this work}
\affiliation{Howard P. Isermann Department of Chemical and Biological Engineering and  Center for Biotechnology and Interdisciplinary Studies, Rensselaer Polytechnic Institute, Troy, NY 12180}
\affiliation{Late Stage Process Development, Genzyme - A Sanofi Company, Framingham, MA 01701}
\author{Sumanth N. Jamadagni}
\thanks{RG and SNJ contributed equally to this work}
\affiliation{Howard P. Isermann Department of Chemical and Biological Engineering and  Center for Biotechnology and Interdisciplinary Studies, Rensselaer Polytechnic Institute, Troy, NY 12180}
\affiliation{Beckett Ridge Technical Center, The Procter and Gamble Company, West Chester, OH 45069}
\author{Vasudevan Venkateshwaran}
\author{Shekhar Garde}
\affiliation{Howard P. Isermann Department of Chemical and Biological Engineering and  Center for Biotechnology and Interdisciplinary Studies, Rensselaer Polytechnic Institute, Troy, NY 12180}

\begin{abstract}
We use molecular simulations to demonstrate the connection between transverse water-water correlations and wetting phenomena for a range of hydrophobic to hydrophilic solid surfaces.Near superhydrophobic surfaces, the correlations are long ranged, system spanning, and are well described by the capillary wave theory. With increasing surface-water attractions, the correlations are quenched. At the critical attraction at which long range correlations disappear, the density profile normal to the surface changes from sigmoidal to layered, and the fluid begins to wet the surface. This behavior is displayed by both water and a Lennard-Jones fluid, highlighting the universality of the underlying physics. 
\end{abstract}


\maketitle

Hydrophobicity, at the macroscale, is reflected in the beading up of water into droplets on a surface. Correspondingly, contact angle, interfacial tension, or the work of adhesion characterize hydrophobicity. Recent theoretical and simulation work has shown that water density fluctuations near a surface provide a robust signature of its hydrophobicity - fluctuations are enhanced near a hydrophobic surface and quenched near hydrophilic ones \cite{GodawatJG09, SarupriaG09, PatelVC2010}. In contrast to fluctuation based measures, one does not generally expect the average water density near a realistic surface to correlate with the wetting properties of that surface \cite{GodawatJG09}. The origin of this failure is hidden in the failure of Weeks-Chandler-Andersen (WCA) based perturbation ideas \cite{WCA} to describe liquid structure near in inhomogeneous interfacial systems (see the ``Exceptions and Qualifications'' section of \cite{WCAReview}). Water (or any other liquid with cohesive interactions) dewets from a repulsive hard wall, creating a vapor-liquid interface with sigmoidal variation of density perpendicular to the wall \cite{Stillinger1973}. However, unlike that in the bulk liquid, the density of the fluid near the wall is highly sensitive to the wall-fluid interactions. Thus, even small attractive interactions ({\it e.g.,} weak van der Waals interactions of water with teflon) are sufficient to increase the local density to bulk-like values, thus masking the underlying dewetting transition. 

\begin{figure}[b]
\centering
\includegraphics[scale=1.0]{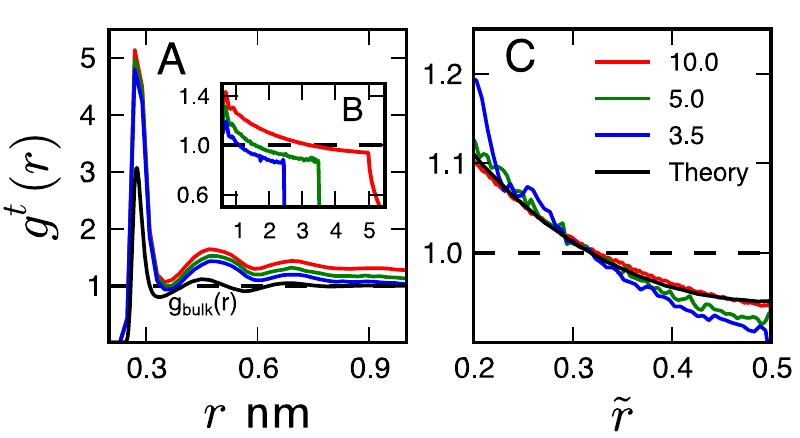}
\caption{(A) Short and (B) long range parts of transverse oxygen-oxygen pair correlations, $g^t(r)$, at the vapor liquid interface of water obtained from MD simulations in 3D periodic boxes of $L\times L$ cross-section, with $L=$ 3.5, 5, and 10 nm. (C) Long range correlations, $g^t(\tilde r)$, as a function of scaled separation $\tilde r=r/L$. $g^t(\tilde r)$ predicted using the capillary wave theory within a multiplicative constant is also shown. Correlations were measured in a slab centered at the half-density plane bounded by planes with average water density of $0.4\rho_{b}$ and $0.6\rho_{b}$, where $\rho_{b}$ is the bulk water density.}
\label{fig:lvgr}
\end{figure}

Density fluctuations are quantified by $P_v(N)$, the probability of finding $N$ solvent molecules in an observation volume $v$ of interest.  Enhanced fluctuations near solvophobic surfaces are reflected in the higher variance of $P_v(N)$ compared to that in bulk solvent \cite{GodawatJG09}, and also in the low-$N$ fat tails of $P_v(N)$, with the fatness of the tail suggestive of the solvent being pushed near its liquid to vapor phase transition by the surface \cite{PatelVC2010, PatelVJHCG2012}. Because the moments of $P_v(N)$ are related to $N$-particle correlations in the fluid, we expect a signature of solvophobicity to be also present in fluid-fluid correlations near the interface.  

Here we focus on the two-particle water-water correlations near surfaces that span a range of hydrophobicity, and show that these correlations serve as a molecular signature of hydrophobicity. For example, near a VL interface or a superhydrophobic surface, water-water correlations are long ranged, system size dependent, and governed by capillary fluctuations. 
With increasing surface hydrophilicity, the correlations are quenched, become short ranged and system size independent. We show that the contact angle as well as the correlation length scale with surface-water attractions. These observations are not unique to water, but also apply to a Lennard-Jones (LJ) fluid near solid surfaces, highlighting the universality of the underlying physics. 

Figure \ref{fig:lvgr} shows the short (panel A) and long (panel B) range parts of transverse ({\it i.e.}, parallel to the interface) pair correlations, $g^t(r)$, between water molecules located at the VL interface. The radial distribution function in bulk water, $g_{bulk}(r)$, is also shown for reference. Consistent with molecular packing and the tetrahedral hydrogen bonded network of water molecules, $g_{bulk}(r)$, shows short-range correlations with first and second peaks at 0.28 and 0.45 nm, respectively. Near the VL interface, the short range correlations are enhanced, consistent with the stronger hydrogen bonding near a low dielectric interface \cite{PatelNG2003}. The location of the second peak is shifted somewhat outward, consistent with the quasi-2D nature of packing and interactions in the interfacial slab \cite{RaschkeTL2001}. 

More importantly, in all systems studied here, we observe system spanning long range correlations. Both one-particle density perpendicular to the interface and the transverse correlations are governed by a balance of thermal fluctuations that deform the interface and surface tension that places an energetic penalty on such deformations. The capillary wave theory (CWT) framework of Buff, Lovett, and Stillinger describes this balance quantitatively by decomposing the instantaneous surface configuration into harmonic waves of various wavelengths that are thermally excited and damped by surface tension \cite{BuffLS65, RowlinsonWidom,Davis77,Weeks77,Evans79}. Because the longest wavelength in a given simulation is bounded by the size of the box, $g^t(r)$ is system size dependent. $g^t(r)$ from different systems, however, show a universal behavior when scaled by the box length, and is described well by the CWT within a multiplicative constant as shown in Figure \ref{fig:lvgr}C \cite{VakninBT08,KalosPR77}. Similar long range correlations were observed in simulations of the VL interface of a Lennard-Jones (LJ) liquid \cite{Wertheim76,SidesGL99} (see Supporting Information for mathematical details and the LJ data).
 
\begin{figure}[h]
\centering
\includegraphics[scale=1.00]{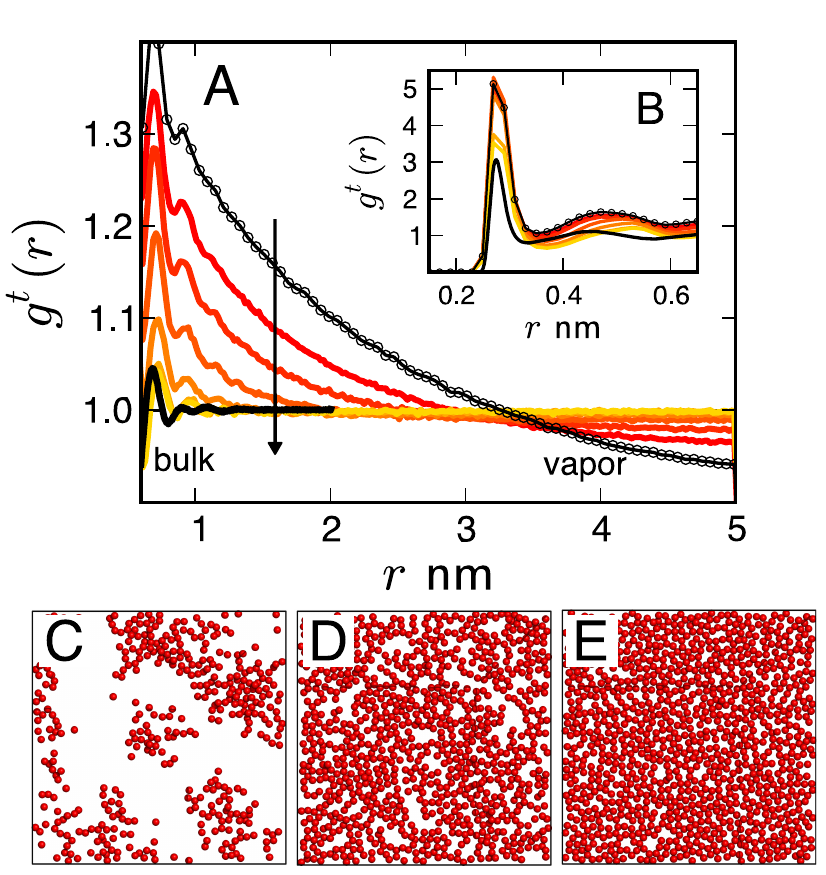}
\caption{(A) Long range and (B) short range parts of the transverse oxygen-oxygen pair correlations, $g^t(r)$, near a solid-water interface obtained from MD simulations of a system with $L=10$ nm (Some data taken from Figure 5d of \cite{JamadagniGG2011}. The surface interacts with water molecules via the 9-3 potential: ${U(z) = \frac{2\pi}{3} \rho _s \epsilon_{sf} \sigma _{sf}^3 \left[ \frac{2}{15} \left( \frac{\sigma _{sf}}{z} \right)^9- \left (\frac{\sigma _{sf}}{z}\right)^3 \right] }$, where $\epsilon_{sf}$ is the strength of attraction and $\rho_{s}$ is the number density of surface atoms per unit area. We use $\sigma_{sf} = 0.355$ nm and $\rho_{s} = 35$ nm$^{-2}$. Six $g^t(r)$ profiles for $\epsilon_{sf}$ = 0.1, 0.35, 0.5, 0.75, 1.0, and 1.5 kJ/mol are shown (red to yellow, see arrow). The correlations in bulk (black line) at the VL interface (circles) are shown for reference. Simulation snapshots of the interfacial slab with (C) $\epsilon_{sf}$ = 0.05 kJ/mol (vapor-liquid-like, $\theta \approx 180^{\circ}$), (D) $\epsilon_{sf}$ = 0.80 kJ/mol (alkane-like, $\theta \approx 120^{\circ}$) and (E) $\epsilon_{sf}$ = 2.50 kJ/mol (hydrophilic, $\theta \approx 0^{\circ}$) are shown. Only water oxygens (red) are shown for clarity.}
\label{fig:slgr}
\end{figure}

How does a fluid respond to the presence of a surface? Stillinger showed that liquids dewet from a purely repulsive surface, {\it e.g.,} a hard wall, and form a vapor-liquid-like interface, the exact location of which is determined by the system pressure and the entropic repulsion between the soft fluctuating interface and the hard wall \cite{Stillinger1973}. How does increasing the hydrophilicity of a surface affect $g^{t}(r)$ in its vicinity? Figure \ref{fig:slgr} shows $g^t(r)$ for water molecules near flat surfaces interacting with water with an attractive 9-3 potential (see caption), with surface-water attractions sampling a broad range of hydrophobicity/philicity. The use of a 9-3 potential, instead of an explicit atomic surface, allows us to isolate the effect of attractions on liquid structure, and avoids spurious correlations that arise from surface topography. 

At small $\epsilon_{sf}$, $g^{t}(r)$ near the solid surface is similar to that at a VL interface -- the short range ($r<0.8$ nm) correlations are enhanced compared to that in bulk water, and the long range correlations span the entire system. With increasing $\epsilon_{sf}$, both the short and long range correlations are quenched, albeit differently. The short-range correlations are not significantly affected as $\epsilon_{sf}$ is increased from 0.05 to 0.35 kJ/mol, but there is a dramatic quenching of the long range correlations. It is only when $\epsilon_{sf}\approx 0.5$ kJ/mol, that the short range part ({\it e.g.,} the first peak) is affected significantly, but is never truly bulk-like even for $\epsilon_{sf}=1.5$ kJ/mol. 

The following physical picture emerges from these observations. The interface of water with an extended hydrophobic surface is soft, vapor-liquid-like, accommodates significant fluctuations, with water molecules at the interface displaying long-range transverse correlations. As the surface becomes more hydrophilic, long-wavelength fluctuations incur significant energetic cost and are gradually quenched, leading to a decrease in the correlation length. In absence of long-range transverse correlations near solvophilic surfaces, the packing of solvent molecules (and their relative orientations and local hydrogen bonding, in case of water) determine the precise nature of the short range correlations. The packing and relative orientations of molecules in the interfacial layer that wets the surface never truly become bulk-like, as a result of quasi-2D (interfacial) vs 3D (bulk) spatial contexts for those molecules. This behavior is not unique to water, as shown by results for Lennard-Jones liquid near the surface (see Supplementary Information), suggesting the generality of the physical picture presented here.

\begin{figure}[h]
\centering
\includegraphics[scale=0.85]{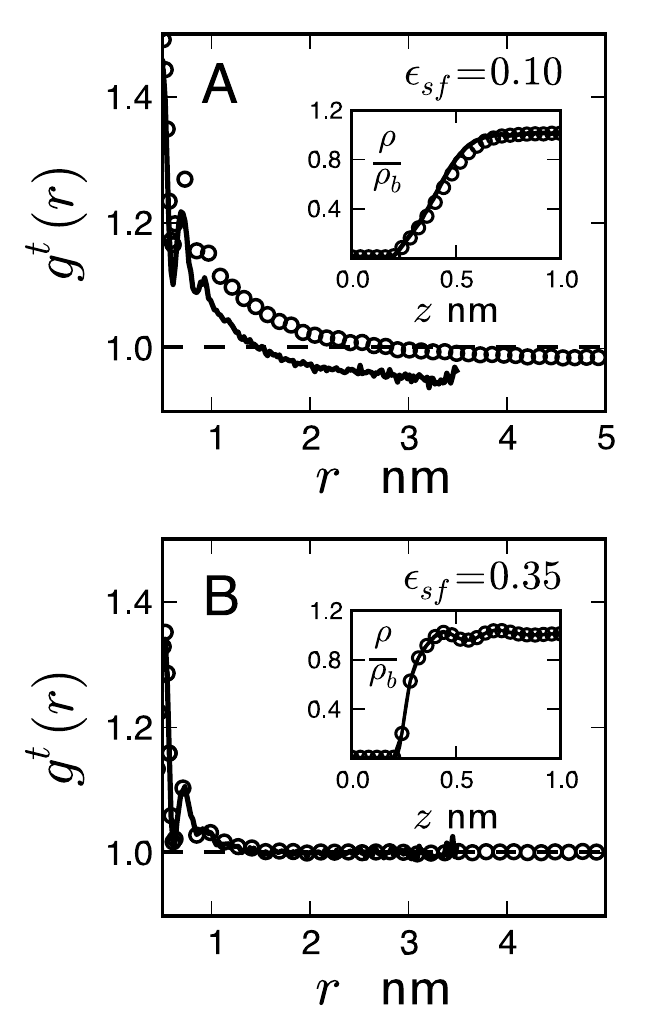}
\caption{Transverse oxygen-oxygen correlation function, $g^t(r)$, near 9-3 surfaces with (A) $\epsilon_{sf}=0.1$ kJ/mol, and (B) $\epsilon_{sf}=0.35$ kJ/mol for two systems with different sizes -- $L=5$ nm (line), $L=10$ nm (open circles). The insets show the normalized one-particle density profile in the direction perpendicular to the surface.}
\label{fig:systemsize}
\end{figure}

Snapshots of water molecules in the interfacial region near surfaces with varying surface-water attractions shown in Figure \ref{fig:slgr} support the above physical picture. At the most hydrophobic surface, $\epsilon_{sf}=0.05$ kJ/mol (water droplet contact angle, $\theta\approx 180^\circ$), the aqueous interface is similar to that near a hard-wall or the VL interface. The average density in the interfacial region is about half of that in the bulk. However, water cannot exist as a homogeneous fluid at such a density, and phase separates into small microscopic clusters, as reflected in the regions of sharply varying densities of $\sim\rho_{b}$ and $\sim$0 in the plane (Figure \ref{fig:slgr}C). This microscopic phase separation of water between the liquid and vapor phases is consistent with the observed long range correlations near this surface. 

At intermediate $\epsilon_{sf}$ values ({\it e.g.,} $\epsilon_{sf}=0.8$ kJ/mol, $\theta\approx 120^\circ$), typical of solid alkane-water interfaces, we observe a largely liquid-like region at the interface, with many small cavities opening up at various locations transiently (Figure \ref{fig:slgr}D). The absence of large vapor-like voids is consistent with the quenching of the long range correlations. Further, bulk-like water density near such a hydrophobic surface is also consistent with the observations of Godawat et al. \cite{GodawatJG09}, indicating the masking of the underlying dewetting transition. Indeed, at such hydrophobic interfaces, water density fluctuations display fat low-$N$ tails in $P_v(N)$ (see Supporting Information), similar to that observed by Patel et al. \cite{PatelVC2010,PatelVJHCG2012}, revealing their vapor-liquid-like nature. Further increase in $\epsilon_{sf}$ leads to a complete wetting ($\theta\approx 0$) of the surface, with the interfacial water layer being similar to a dense fluid with minimal density fluctuations and no observable vapor-like cavities (Figure \ref{fig:slgr}E).

Data in Figure \ref{fig:systemsize} underscores the connections between the range of transverse correlations, system size dependence of simulation results, and the corresponding qualitative changes in the nature of one-particle density of water normal to the surface. For $\epsilon_{sf}=0.1$ kJ/mol, the surface is highly hydrophobic ($\theta\approx 180^\circ$), $g^t(r)$ is long ranged, and spans the system for both $L=$5 and $L=$10 nm systems. That is, $g^t(r)$ is system size dependent. The one-particle density profile of water normal to this surface is sigmoidal, similar to that near a VL interface, and displays a weak system size dependence, with the interface for $L=10$ nm system being broader than for $L=5$ nm. CWT predicts a logarithmic increase of the interface width with system size, and our results are qualitatively consistent with that prediction.

Increasing the surface-water attractions modestly to $\epsilon_{sf}=0.35$ kJ/mol decreases the contact angle somewhat (to $ \approx 160^\circ$), suggesting that this surface is still very hydrophobic. Yet, this small change in $\epsilon_{sf}$ leads to several important qualitative changes in the nature of fluid structure at the interface. First, the transverse correlations are no longer system spanning, making $g^t(r)$ independent of system size, at least for $L\geq 5$ nm. Also, the one particle density profile is no longer sigmoidal, instead layering of water is observed. This high sensitivity of the local water density to $\epsilon_{sf}$ near a hydrophobic surface is contrary to the WCA picture of the effects of attractions on fluid structure in bulk homogeneous liquids. In fact, Weeks, Chandler, and Andersen had anticipated such sensitivity (see ``Exceptions and Qualifications'' section in \cite{WCAReview}). They emphasize that in situations where the compressibility is sufficiently high to allow for long wavelength fluctuations ({\it i.e.,} when repulsive cores are not sufficiently effective in screening the interparticle correlations), the classic WCA picture will break down. 

\begin{figure}[h]
\centering
\includegraphics[scale=0.85]{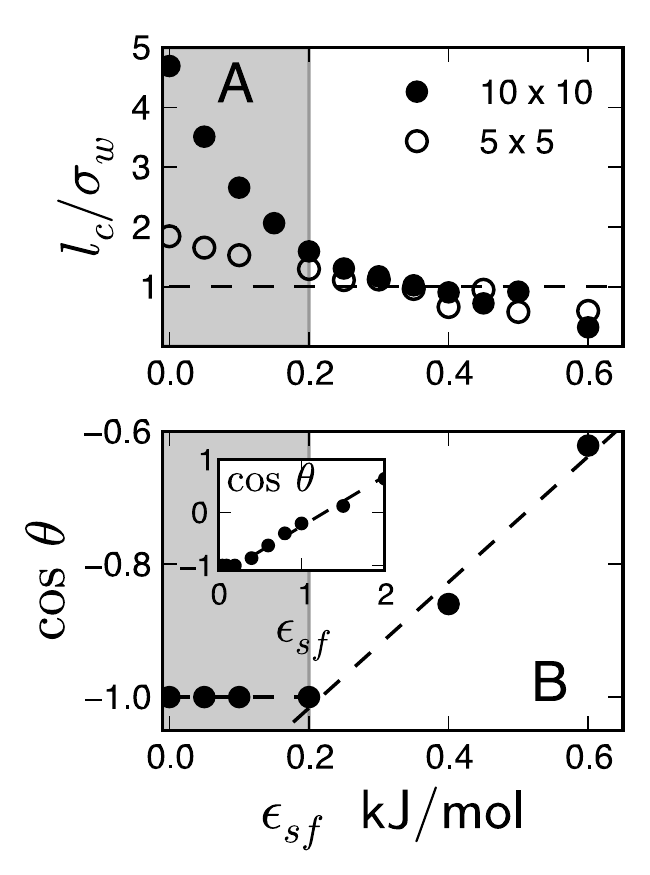}
\caption{ (A) The correlation length, $l_c$, normalized by the size of a water molecule, $\sigma_w = 0.30$ nm, as a function of the surface-water attraction, $\epsilon_{sf}$. $l_c$ is obtained by exponential fits to normalized transverse correlation function at separations $r>3.5\sigma_w$. That is, $\tilde{g}^t(r)\sim \exp\left(-r/l_c\right)$, where $\tilde{g}^{t}(r) = \frac{g^{t}(r) - g^{t}(L/2)}{g^{t}(3.5\sigma_{w}) - g^{t}(L/2)}$. Data are shown for two system sizes, $L=5$ and 10 nm.  (B) Cosine of the contact angle $\theta$ for water droplets containing 2100 molecules placed on the 9-3 surface as a function of $\epsilon_{sf}$. The inset shows the same data for a larger range of $\epsilon_{sf}$ values.}\label{fig:lctheta}
\end{figure}

Figure \ref{fig:lctheta}A shows that for small $\epsilon_{sf}$, the transverse correlation length normalized by the size of a water molecule, $l_c/\sigma_w$, is large and depends on the size of the system. Increasing the surface-water attractions reduces $l_c/\sigma_w$ monotonically, and for values of $\epsilon_{sf}$ larger than $\epsilon_{sf}^{*} \approx 0.2$ kJ/mol, the correlations become short ranged, $l_c/\sigma_w \approx 1$, and system size independent. A similar reduction in the correlation length at fluid interfaces by external fields ({\it e.g.}, gravitational field) has been discussed by Knackstedt and Robert \cite{KnackstedtR88}.

Figure \ref{fig:lctheta}B points to a connection of these microscopic results with macroscopic characterization of hydrophobicity in terms of a water droplet contact angle, $\theta$. We expect a purely repulsive surface to be superhydrophobic with $\cos \theta = -1$. Interestingly, we find that for $\epsilon_{sf} < \epsilon_{sf}^{*}$, the surface-water attraction is not sufficiently strong to compete with thermal capillary fluctuations and the surface remains superhydrophobic. For $\epsilon_{sf} > \epsilon_{sf}^{*}$, however, we observe a linear variation of $\cos \theta$ with $\epsilon_{sf}$ similar to what was observed by Sendner et al. \cite{SendnerHBN2009}. Assuming that the solid-liquid interfacial tension, $\gamma_{sl}$, will decrease approximately linearly with the strength of attractive interactions \cite{Vazquez2009}, such linear variation in $\cos \theta$ with $\epsilon_{sf}$ is expected with the slope $m_{\theta} \approx -(\partial \gamma_{sl}/\partial \epsilon_{sf})/\gamma_{lv}$, where $\gamma_{lv}$ is the liquid-vapor interfacial tension. The exact value of $\epsilon^{*}$ is difficult to pinpoint using the methods employed here. Nevertheless, the consistency between results in Figures \ref{fig:lctheta}A and B supports the underlying physical picture.

\begin{figure}
\centering
\includegraphics[scale=1.0]{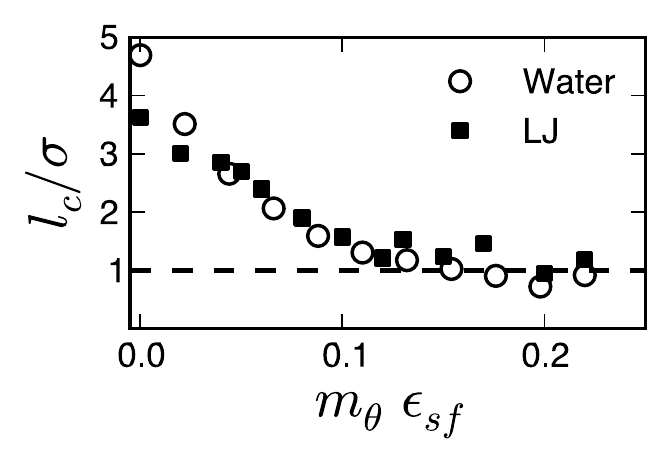}
\caption{Dependence of the normalized correlation length, $l_{c}/\sigma$, on the scaled surface-fluid attraction, $m_{\theta}\epsilon_{sf}$ for water and LJ fluid.}\label{fig:lcscaling}
\end{figure}

The dependence of $l_{c}/\sigma_{LJ}$ and of $\cos \theta$ on $\epsilon_{sf}$ for a LJ fluid is qualitatively similar to that shown by water (see Supplementary Information), suggesting a universal scaling of the $l_{c}/\sigma$ for both fluids. 
Data in Figure \ref{fig:lcscaling} showing the variation of $l_{c}/\sigma$  with $m_{\theta}\epsilon_{sf}$ for both water and LJ fluid confirms this suggestion. The small differences in the behavior of $l_{c}/\sigma$ at very low $\epsilon_{sf}$  result from the differences in system sizes in units of $\sigma$ ($\sigma_w$=0.30 nm; $\sigma_{LJ}$ = 0.37 nm). This difference results in a larger number of capillary wave harmonics for water for the same system size (10 nm), and consequently, higher $l_{c}/{\sigma_{w}}$. Our results suggest new avenues for experimental investigations of inhomogeneous and confined fluids and add to the growing knowledge about the characterization of hydrophobicity of complex surfaces at the molecular level.

Our results reveal the connections between three seemingly independent characterizations of inhomogeneous fluids -- one particle density, transverse correlations and wetting. We showed that the transition from sigmoidal to layered fluid density profile, quenching of long ranged transverse pair correlations, and the onset of surface wetting are interconnected and occur simultaneously with increasing solvophilicity of the surface. 

\begin{acknowledgments}
This work was supported by the National Science Foundation grant CBET-1159990. We thank Center for Computational Innovations at Rensselaer Polytechnic Institute for high-performance computing resources. 
\end{acknowledgments}


%

\end{document}


\title{Supporting Information for \\ Connecting water correlations, fluctuations, and wetting phenomena at hydrophobic and hydrophilic surfaces} 

\author{Rahul Godawat}
\thanks{RG and SNJ contributed equally to this work}
\affiliation{Howard P. Isermann Department of Chemical and Biological Engineering and  Center for Biotechnology and Interdisciplinary Studies, Rensselaer Polytechnic Institute, Troy, NY 12180}
\affiliation{Late Stage Process Development, Genzyme-A Sanofi Company, Framingham, MA 01701}
\author{Sumanth N. Jamadagni}
\thanks{RG and SNJ contributed equally to this work}
\affiliation{Howard P. Isermann Department of Chemical and Biological Engineering and  Center for Biotechnology and Interdisciplinary Studies, Rensselaer Polytechnic Institute, Troy, NY 12180}
\affiliation{Beckett Ridge Technical Center, The Procter and Gamble Company, West Chester, OH 45069}
\author{Vasudevan Venkateshwaran}
\author{Shekhar Garde}
\affiliation{Howard P. Isermann Department of Chemical and Biological Engineering and  Center for Biotechnology and Interdisciplinary Studies, Rensselaer Polytechnic Institute, Troy, NY 12180}

\maketitle

\section{Methods}
\subsection{System geometry and forcefields} 
Our simulation system consists of a slab of liquid in contact with a vapor layer above it in the $z$ direction, forming a vapor--liquid interface. The bottom layer of the liquid slab is anchored to a surface modeled by a potential, $U(z)$, acting uniformly across the cross-section, given by:

\begin{equation}
{U(z) = \frac{2\pi}{3} \rho _s \epsilon _{sf} \sigma _{sf}^3 \left[ \frac{2}{15} \left( \frac{\sigma _{sf}}{z} \right)^9- \left
(\frac{\sigma _{sf}}{z}\right)^3 \right] } \protect\label{eqn:uz}
\end{equation}

\noindent where $\sigma_{sf}= 0.355$ nm, $\rho_{s}$ (= 35 nm$^{-2}$) is the number density of surface atoms per unit area, and $\epsilon_{sf}$ is the strength of surface-fluid interactions that is varied as described in the main text. 

\begin{figure}[h]
\centering
\includegraphics[scale=0.8]{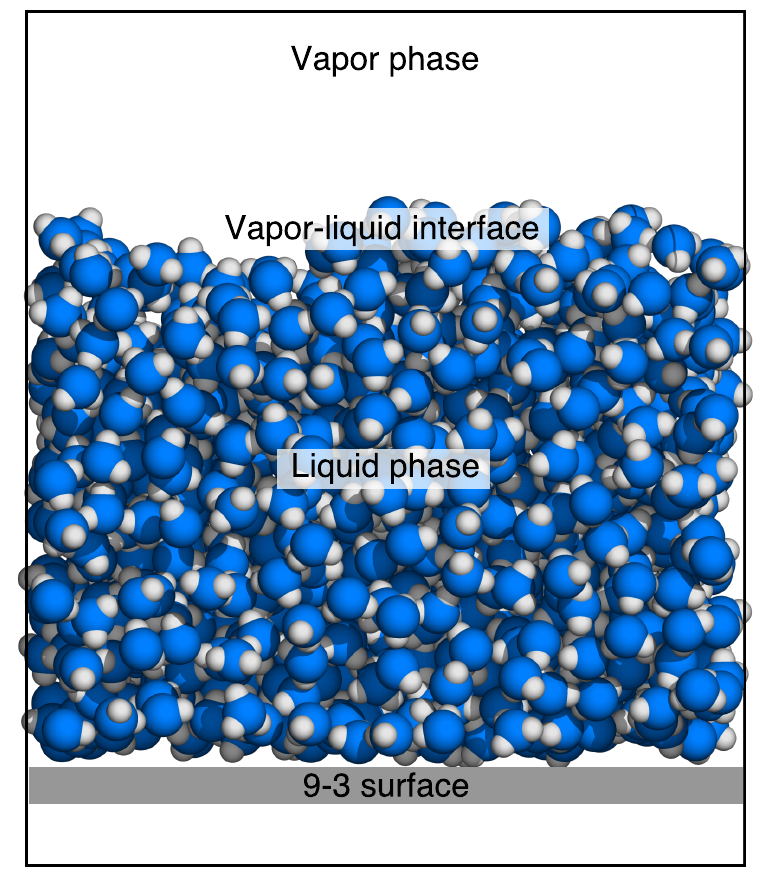}
\caption{Simulation snapshot of a water system. Water molecules are shown in a spacefill representation with oxygen atoms colored blue and hydrogen atoms colored light gray. The 9-3 surface is marked by a gray slab.}\label{systemschematic}
\end{figure}

We study two liquids (i) water represented explicitly using the extended simple point charge (SPC/E) model and (ii) Lennard-Jones (LJ) fluid with the interparticle potential described by: 

\[ U(r) = 4\epsilon_{LJ}\left[ \left( \frac{\sigma_{LJ}}{r} \right)^{12} - \left( \frac{\sigma_{LJ}}{r} \right)^{6} \right] \]

\noindent where $\sigma_{LJ}=0.373$ nm and $\epsilon_{LJ}=1.234$ kJ/mol. We performed simulations of both fluids in systems of various cross-sections and with a range of surface-fluid interactions. The details of the cross-sectional area and number of molecules included in each simulation are given in Table. \ref{tab:systems}. 

\begin{table}[h]
\caption{System details}\label{tab:systems}
\centering
\begin{tabular}{L{1.5cm}C{4cm}C{2cm}}
\hline
Fluid & Box cross-section (nm$^{2}$) & $N$  \\
\hline
Water & 3.5 $\times$ 3.5  & 1000  \\
      & 5 $\times$ 5  & 2447  \\
      & 10 $\times$ 10  & 8287  \\
\hline
LJ    &  5 $\times$  5  & 2055  \\
      & 10 $\times$ 10  & 9849  \\
      & 15 $\times$ 15  & 25836 \\
      & 20 $\times$ 20  & 52436 \\
\hline
\end{tabular} 
\end{table}

\subsection{Simulation details} 
Simulations were performed in the canonical ensemble (NVT) using the molecular dynamics package GROMACS \cite{Gromacs}. The Leapfrog algorithm with a time-step of 2 fs was used to integrate the equations of motion with bond constraints on water molecules imposed using the SHAKE algorithm \cite{SHAKE}. All simulations of water were carried out at a temperature of 300 K and of the LJ fluid at 104 K. The temperature was maintained using a Nose-Hoover thermostat \cite{NoseHoover1, NoseHoover2}. Electrostatic interactions were calculated using the particle mesh Ewald (PME) algorithm with a grid spacing of 0.12 nm and a real-space cut-off of 1.0 nm \cite{PME}. Lennard-Jones interactions were also truncated at 1.0 nm. Each simulation was equilibrated for 500 ps followed by a production of run of more than 10 ns. Configurations were stored every picosecond for further analysis.

\section{Density profiles at the solid-liquid interface}
As described in the main text, we studied the solvation of 9-3 surfaces by water and a LJ fluid. Whereas the main manuscript focuses on capillary fluctuations and transverse correlations, for completeness here we report the one particle density profile of these two fluids in the direction perpendicular to the surface as a function of surface-fluid attractions. Because we have a well-defined flat mathematical surface, the local density of fluid increases monotonically with increasing attractions. 

\begin{figure}[h]
\centering
\includegraphics[scale=0.8]{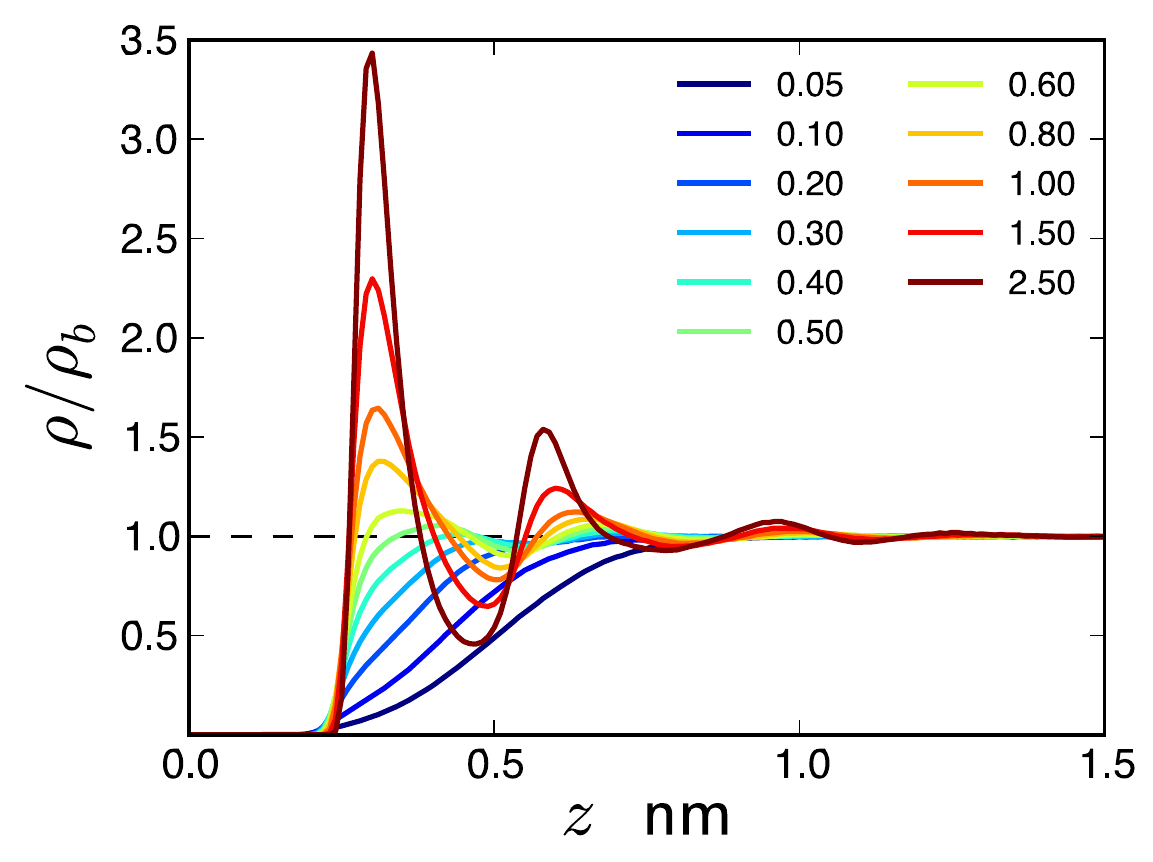}
\caption{Density profile of water perpendicular to the interfacial plane for various surface-water attractions. The solid surface is located at $z=0$.}\label{waterdensity}
\end{figure}

\begin{figure}[h]
\centering
\includegraphics[scale=0.8]{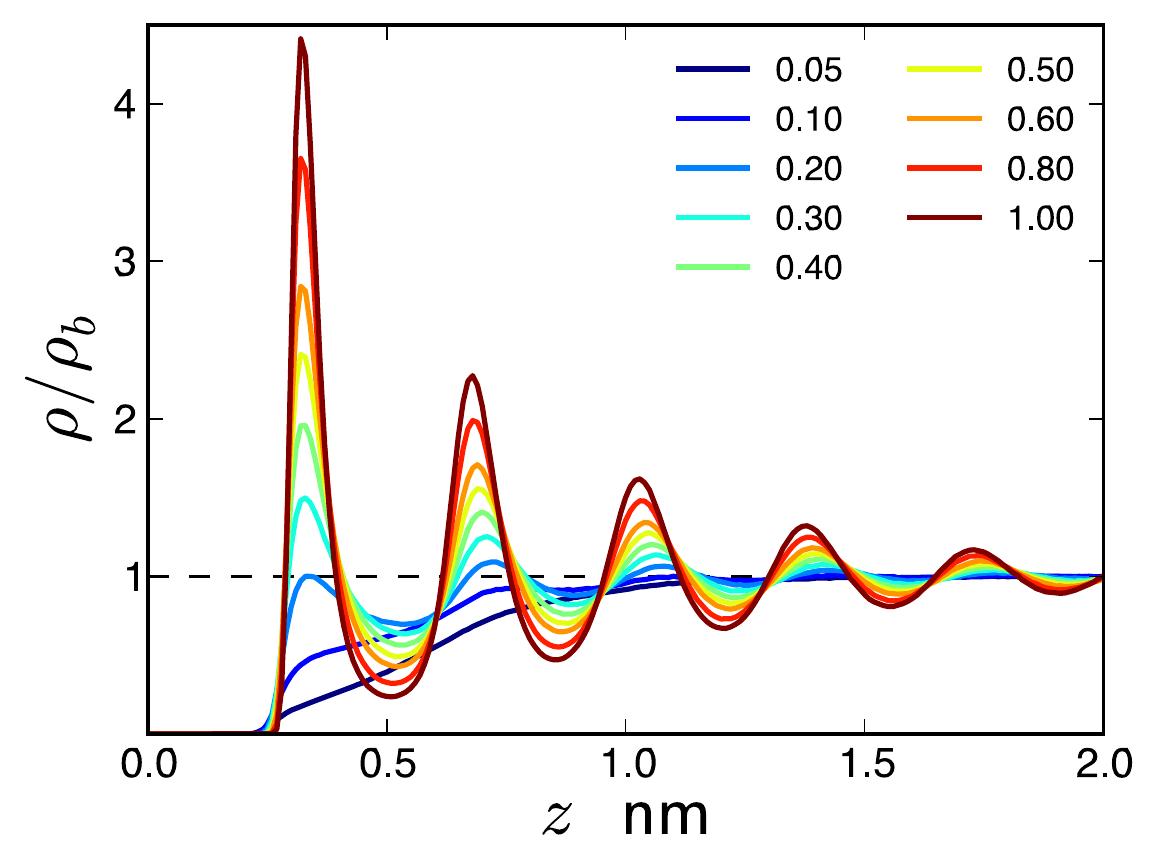}
\caption{Density profile of Lennard-Jones liquid perpendicular to the interfacial plane for various surface-fluid attractions. The solid surface is located at $z=0$.}\label{ljdensity}
\end{figure}

\section{Capillary Wave Theory for the liquid-vapor interface}
The capillary wave theory (CWT) for liquid vapor interfaces represents the instantaneous surface configuration in terms of surface harmonic waves:

\begin{equation}
h(x,y) = \displaystyle \sum_{k} A(k) e^{i\vec{k}.\vec{r}}
\end{equation}

The probability of observing a particular set of $A(k)$ values is proportional to the Boltzmann factor $\exp \left( -W/\kt \right)$, where $W$ is the work done against the interfacial tension and any other external forces (such as due to a surface). For a liquid-vapor interface under no external field \citep{BuffLS65,KalosPR77}, 

\begin{equation}
\ens{\left| A(k)\right|^{2}} \propto \frac{\kt}{\gamma k^{2}}
\end{equation} 

\noindent where $\gamma$ is the liquid-vapor interfacial tension. The transverse solvent structure factor, $S^{t}(k) \propto 1/k^2$ at the interface \cite{VakninBT08,KalosPR77}. The transverse correlations, $g^t(r)$ can be obtained from an inverse Fourier transform:

\begin{equation}\protect\label {eqn:lvgtr}
g^{t}(r) - 1 \propto \displaystyle\sum_{k} \frac {|e^{{\it i}{\bf k.r}}|}{k^2}
\end{equation}

\begin{figure}[h]
\centering
\includegraphics[scale=1.0]{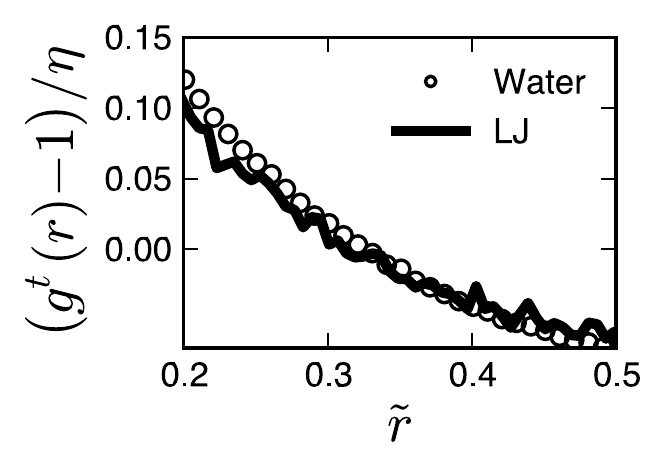}
\caption{Scaling of long-range correlations by the CWT scaling constant $\eta$. }\label{fig:waterljscaling}
\end{figure}

The proportionality constant, $\eta$, in Eqn \ref{eqn:lvgtr} is given by \cite{KalosPR77}:

\begin{equation}\protect\label{eqn:eta}
\eta = \frac{kT}{\gamma}\left[\frac{\rho '(z_o)^2}{\rho(z_o)}\right]
\end{equation}

\noindent where $\rho(z)$ is the density profile along the normal to the interface and $z_o$ is the location of the interface. $\rho(z)$ is frequently represented using \cite{AlejandreTC1995}:

\begin{equation}
\frac{\rho(z)}{\rho_{b}} = 0.5\left[1+\tanh \left(\frac{z-z_o}{d}\right)\right]
\end{equation}

\noindent where $d$ is a measure of the width of the interface and $\rho_{b}$ is the bulk density. Figure \ref{fig:waterljscaling} shows long range transverse correlations in water and LJ fluid scaled by $\eta$.

\section{Transverse correlations and contact angle data for a Lennard-Jones fluid}
\begin{figure}[h]
\centering
\includegraphics[scale=1.0]{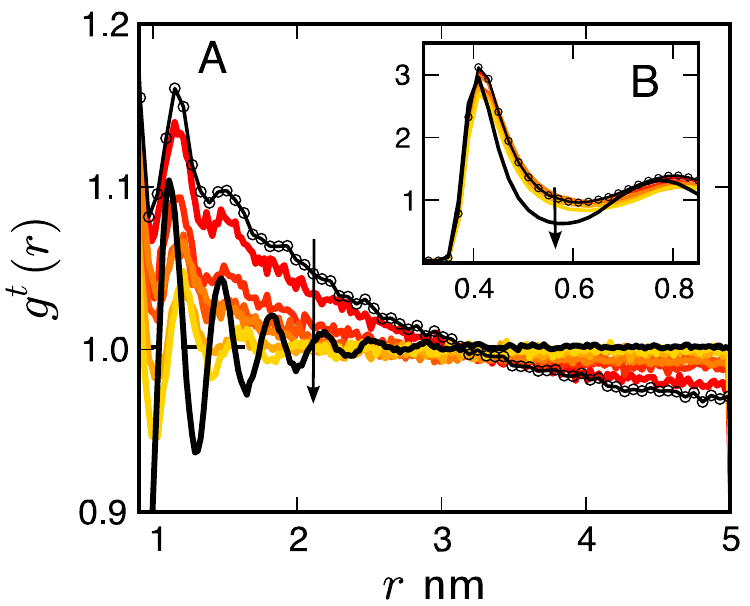}
\caption{Correlations at the solid-LJ fluid interface. Six $g^t(r)$ profiles for $\epsilon_{sf}$ = 0.02, 0.05, 0.08, 0.1, 0.2, and 0.3 kJ/mol are shown (red to yellow, see arrow). The correlations in bulk (black line) at the VL interface (circles) are shown for reference. Direction of arrow indicates increasing surface-fluid attractions}\label{fig:slgrlj}
\end{figure}

\begin{figure}[h]
\centering
\includegraphics[scale=1.0]{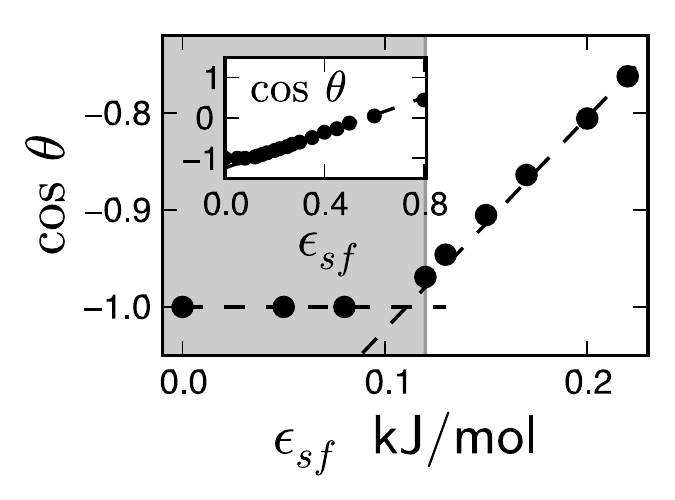}
\caption{Cosine of the contact angle $\theta$ for LJ liquid droplets on the 9-3 surface as a function of $\epsilon_{sf}$.}\label{fig:ljtheta}
\end{figure}

The main text uses data on transverse correlations and contact angles for liquid water. As described in the main text, the underlying physics is general, and trends similar to that in water are observed for a LJ fluid as shown in Figures \ref{fig:slgrlj} and \ref{fig:ljtheta}.

\section{Water density fluctuations at the solid-liquid interface}
Water density fluctuations at the interface are quantified by the probability distribution $P_{v}(N)$, that describes the probability of observing N waters in an observation volume $v$. We employed the INDUS method \cite{PatelVCG2011} to calculate $P_{v}(N)$ in cuboidal observation volumes of dimension 1.5 nm x 1.5 nm x 0.5 nm placed adjacent to the solid surface.

\begin{figure}[h]
\centering
\includegraphics[scale=1.0]{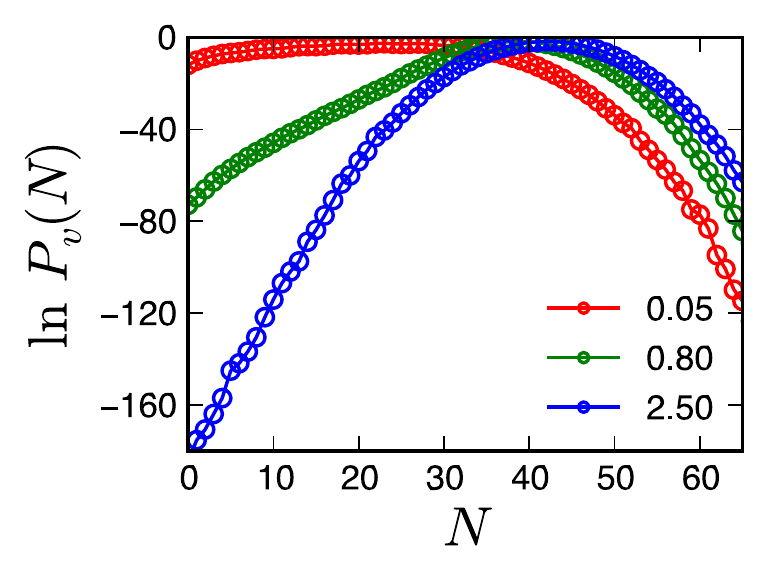}
\caption{Probablity distributions of the number of water oxygen centers in a 1.5 nm x 1.5nm x 0.5 nm volume adjacent to the 9-3 surface calculated using the Indirect Umbrella Sampling (INDUS) method for three different surface-fluid attractions.}\label{fig:indus}
\end{figure}

Figure \ref{fig:indus} shows $\log P_{v}(N)$ for water next to 9-3 surfaces with increasing surface-water attractions. For $\epsilon_{sf} = 2.50$ kJ/mol, the $P_{v}(N)$ is approximately Gaussian and the probability of observing $N=0$ water molecules in the probe volume is negligible. As $\epsilon_{sf}$ is decreased, we observe low-$N$ fat tails in the $\log P_{v}(N)$ distribution characteristic of hydrophobic nature of the surface (See \cite{PatelVC2010} for a detailed discussion). 


%